\documentstyle[psfig,aps,prl,multicol]{revtex}
\begin{document}

\title{Nested quantum search and NP-complete problems}

\author{Nicolas J. Cerf,$^{1,2,3}$ Lov K. Grover,$^4$ 
        and Colin P. Williams$^2$}
\address{$^1$W. K. Kellogg Radiation Laboratory,
California Institute of Technology, Pasadena, California 91125\\
$^2$Information and Computing Technologies Research Section,
Jet Propulsion Laboratory, Pasadena, California 91109\\
$^3$Center for Nonlinear Phenomena and Complex Systems,
Universit\'e Libre de Bruxelles,
1050 Bruxelles, Belgium\\
$^4$3C-404A Bell Laboratories, 600 Mountain Avenue, Murray Hill, 
New Jersey 07974}

\date{June 1998}

\draft
\maketitle

\begin{abstract}
A quantum algorithm is known that solves an unstructured search problem
in a number of iterations of order $\sqrt{d}$, where $d$ is
the dimension of the search space, whereas any classical
algorithm necessarily scales as $O(d)$.
It is shown here that an improved quantum search algorithm can be
devised that exploits the structure of a tree search problem by 
{\em nesting} this standard search algorithm.
The number of iterations required to find the solution
of an average instance of a constraint satisfaction problem
scales as $\sqrt{d^\alpha}$, with a constant $\alpha<1$ depending on the
nesting depth and the problem considered. 
When applying a single nesting level
to a problem with constraints of size~2 such as the graph coloring problem,
this constant $\alpha$ is estimated to be around~0.62 
for average instances of maximum difficulty.
This corresponds to a square-root speedup
over a classical nested search algorithm, of which our presented algorithm
is the quantum counterpart.
\end{abstract}
\pacs{PACS numbers: 03.67.Lx, 89.70.+c
      \hfill KRL preprint MAP-225}

\bigskip


\section{Introduction}

Over the past decade there has been steady progress in the development of
quantum algorithms. Most attention has focused on
the quantum algorithms for finding the factors of a composite 
integer~\cite{bib_shor0,bib_shor1} 
and for finding an item in an unsorted
database~\cite{bib_grover0,bib_grover1}. These successes have
inspired several researchers to look for quantum algorithms that can solve
other challenging problems, such as decision problems~\cite{bib_farhi}
or combinatorial search problems~\cite{bib_hogg}, 
more efficiently than their classical counterparts.
\par

The class of NP-complete problems includes
the most common computational problems encountered
in practice~\cite{bib_garey}. In particular, it includes
scheduling, planning, combinatorial optimization, theorem proving,
propositional satisfiability and graph coloring. 
In addition to their ubiquity, NP-complete problems share a fortuitous
kinship: any NP-complete problem can be mapped into any other NP-complete
problem using only polynomial resources~\cite{bib_garey}. Thus, any quantum
algorithm that speeds up the solution of one NP-complete problem
immediately leads to equally fast quantum algorithms
for all NP-complete problems (up to the polynomial cost of translation).
Unfortunately, NP-complete problems appear to be even harder
than the integer factorization problem.  Whereas, classically, the
best known algorithm for the
latter problem scales only sub-exponentially~\cite{bib_lenstra}, NP-complete
problems are widely believed to be exponential~\cite{bib_garey}.
Thus, the demonstration that Shor's quantum
algorithm~\cite{bib_shor0,bib_shor1} can factor an integer in a time that is
bounded by a polynomial in the ``size'' of the integer (i.e., the number of
bits needed to represent that integer), while remarkable,
does not lead to a polynomial-time quantum algorithm
for NP-complete problems, the existence of which being considered as highly
improbable~\cite{bib_bbbv}. Moreover, it has proven to be very difficult
to adapt Shor's algorithm to other computational applications.
\par

By contrast, the unstructured quantum search 
algorithm~\cite{bib_grover0,bib_grover1} can be adapted quite readily to the
service of solving NP-complete problems.  As a candidate
solution to an NP-complete problem can be tested for correctness in
polynomial time, one simply has to create a
``database'' consisting of all possible candidate solutions
and apply the unstructured quantum search algorithm.
Unfortunately, the speedup afforded
by this algorithm is only $O(\sqrt{N})$ where $N$ is the number 
of candidate solutions to be tested. For
a typical NP-complete problem in which one has to find an assignment of 
one of $b$ values to each of $\mu$ variables,  
the number of candidate solutions, $b^\mu$,
grows exponentially with $\mu$.  A classical algorithm would therefore take a
time $O(b^\mu)$ to find the solution whereas the unstructured quantum
search algorithm would take $O(b^{\mu/2})$. 
Although this is an impressive speedup,
one would still like to do better.
\par

While there is now good evidence that for unstructured
problems, the quantum search algorithm is 
optimal~\cite{bib_bbbv,bib_boyer,bib_zalka},
these results have raised the question of whether faster quantum
search algorithms might be found for problems that possess 
{\em structure}~\cite{bib_hogg,bib_hogg2,bib_farhi2,bib_grover-st}.
It so happens that NP-complete problems have such structure
in the sense that one can often build up
complete solutions (i.e., value assignments for all the variables) by
extending {\em partial} solutions (i.e., value assignments for a
subset of the variables). Thus, rather than performing an unstructured
quantum search amongst {\em all} the candidate solutions, in an
NP-complete problem, we can perform a quantum search amongst the 
{\em partial} solutions in order to narrow 
the subsequent quantum search amongst their descendants. 
This is the approach presented in this paper and which
allows us to find a solution to an NP-complete problem in a time that
grows, on average, as $O(b^{\alpha\mu/2})$ for the hardest problems,
where $\alpha<1$ is a constant depending on the problem instance considered.
\par

Our improved quantum search algorithm works by {\em nesting}
one quantum search within another. Specifically, by performing a quantum
search at a carefully selected level in the tree of partial solutions,
we can narrow the effective quantum search amongst the candidate solutions
so that the net computational cost is minimized. 
The resulting algorithm is the quantum counterpart 
of a {\em classical} nested search algorithm which scales
as $O(b^{\alpha\mu})$, giving a square root speedup overall.
The nested search procedure mentioned here
corresponds to a {\em single} level of (classical or quantum) nesting,
but it can be extended easily to several nesting levels.
Thus, our result suggests a systematic technique for
translating a nested classical search algorithm into
a quantum one, giving rise a square-root speedup, which
can be useful to accelerate {\em efficient} classical algorithms
(rather than a simple exhaustive search, of no practical use).
We believe this technique is applicable in all structured quantum searches.
\par

The outline of the paper is as follows. 
Section~\ref{sect_nested_class} introduces a simple classical tree
search algorithm that exploits problem structure to localize the search for
solutions amongst the candidates.  This is not intended to be a
sophisticated classical tree search algorithm, but rather is aimed at
providing a baseline against which our quantum algorithm can be compared.
In Section~\ref{sect_unstruc_qu}, we outline the standard unstructured
quantum search algorithm~\cite{bib_grover0,bib_grover1}.
We focus especially on the algorithm based on an arbitrary
unitary search operator~\cite{bib_grover2}, as this is a key
for implementing quantum nesting. Finally, Section~\ref{sect_nested_quant}
describes the quantum tree search algorithm based on nesting,
which is a direct quantum analog of the classical search algorithm 
appearing in Section~\ref{sect_nested_class}.
The quantum search algorithm with several levels of nesting
is also briefly discussed.
We conclude by showing that the expected time to find a solution grows as
$O(b^{\alpha\mu/2})$, that is, as the square root of the classical time
for problem instances in the hard region. The constant
$\alpha$, depending on the problem considered, is shown to decrease
with an increasing nesting depth (i.e., an increasing number
of nesting levels).

\section{Nested classical search on structured problems} 
\label{sect_nested_class}

\subsection{Structured search in trees}

Many hard computational problems, such as propositional satisfiability,
graph coloring, scheduling, planning, and combinatorial optimization, can
be regarded as examples of so-called ``constraint satisfaction problems''.
Constraint satisfaction problems consist of a set
of variables, each having a finite set of domain values, together with a
set of logical relations (or ``constraints'') amongst the variables 
that are required to hold simultaneously.  A solution is defined by a
complete set of variable/value assignments such that every variable has
some value, no variable is assigned conflicting values, and all the
constraints are satisfied.
\par

In such constraint satisfaction problems, there is often a degree of
commonality between different non-solutions.  One typically finds,
for example, that certain combinations of assignments of values
to a subset of the variables are inconsistent (i.e., violate one or
more of the constraints) and cannot, therefore, 
participate in any solution. These commonalities (several
non-solutions sharing the same ancestor that is inconsistent)
can be exploited to focus the search for a solution.  
Thus, a classical {\em structured} search algorithm can find 
a solution to a constraint satisfaction problem in fewer steps
than that required by a unstructured search by avoiding 
regions of the search space that can be guaranteed to be devoid of solutions.
Before investigating whether the problem structure
can be exploited in a quantum search (see Sec.~\ref{sect_nested_quant}),
we need to understand the circumstances under
which knowledge of problem structure has the potential to be useful,
classically. 
The key idea is that one can obtain complete solutions to a constraint
satisfaction problem by systematically extending partial solutions, i.e.
variable/value assignments that apply only to a subset of the variables in the
problem.  Not all partial solutions are equally desirable however. A
partial solution is ``good'' if it is consistent with all the constraints
against which it may be tested.  Conversely a partial solution is
``nogood'' if it violates one or more such constraints.  Sophisticated
search algorithms work by incrementally extending good partial solutions
and systematically terminating nogood partial solutions.  This induces a
natural tree-like structure on the search space of partial solutions.
\par

\begin{figure}
\caption{Constraint satisfaction problem 
in which we must find an assignment to 
the $\mu$ variables $x_1,x_2,\cdots x_\mu$. As an example, we picture
the {\em graph coloring problem}, in which we have to assign one of
$b$ possible colors to each node of a graph so that
every pair of nodes that are connected directly
have different colors. The
corresponding search tree is characterized by a depth $\mu$
and a branching ratio $b$. By looking at partial solutions at level $i$ 
in the tree (the search space being of size $b^i$)
and considering only the descendants at level $\mu$
of these partial solutions, one
avoids having to search through the entire space at the bottom
of the tree (of size $b^\mu$). }
\vskip 0.25cm
\centerline{\psfig{file=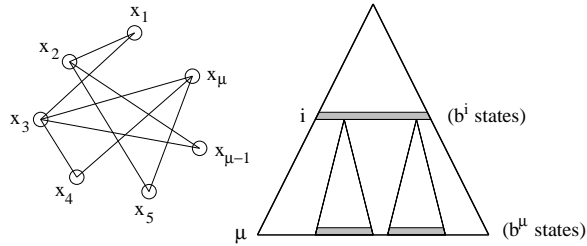,width=3.0in,angle=-90}}
\label{fig_tree}
\vskip -0.25cm
\end{figure}

To give a concrete example of a tree search problem, we consider 
the {\em graph coloring problem} as pictured in Fig.~\ref{fig_tree}.
We have a graph that consists of $\mu$ nodes connected by $e$ edges,
with $0\le e \le \mu(\mu-1)/2$. Each node must be assigned a color 
(out of $b$ possible colors), so that any two nodes connected by
an edge have different colors. More generally, for a constraint satisfaction
problem, we are given a set of $\mu$ variables ($x_1,\cdots, x_\mu$)
to which we must assign a value out of $b$ possible values.
This assignment must satisfy simultaneously a set of constraints, 
each involving $k$ variables. The resulting number of {\em nogood}
ground instances (roughly proportional to the
number of constraints) is denoted by $\xi$.
In the particular case of the graph coloring problem, 
the size of the constraints $k=2$ since each edge imposes
a constraint on the colors assigned to the pair of nodes it connects.
The number of nogood ground instances
$\xi=e b$ because each edge contributes exactly $b$ nogoods and
there are a total of $e$ edges (for each edge, $b$ pairs of
identical colors are forbidden).
\par

The search tree corresponding to this constraint satisfaction problem
is also shown in Fig.~\ref{fig_tree}.
The $i$-th level of the search tree enumerates all possible {\em partial}
solutions involving a specific subset of $i$, out of the total $\mu$,
variables.  The branching ratio in this tree, i.e. the number of children
per node, is equal to $b$, the number 
of domain values of a variable.
For a hard instance of the problem, the number of steps required
to find a good assignment at the bottom of the tree (or decide that
there is no possible assignment satisfying all the constraints)
scales as $b^\mu$, i.e., of the order of 
the entire space of candidate solutions must be explored.
Remarkably, many of the properties of search trees can be understood
without precise knowledge of the constraints. 
Specifically, it has been found empirically
that the difficulty of solving a particular instance 
of a constraint satisfaction problem can be approximately specified
by four parameters: the number of variables,
$\mu$, the number of values per variable, $b$, the number of variables
per constraint, $k$, and the total number of assignments of 
the individual constraints that are 
nogood, $\xi$~\cite{bib_Cheeseman91,bib_Williams92,bib_Williams93}.
\par

Clearly, if $\xi$ is small, there are generally many solutions
satisfying the few constraints, so that the problem is easy to solve. 
Conversely, if $\xi$ is large, the problem is in general overconstrained, 
and it is easy to find that it admits no solution.
The problem is maximally hard in an intermediate range
of values for $\xi$.  In an effort to understand the observed variation in
difficulty across different instances of NP-complete problems
for fixed $\mu$ and $b$, it has been shown
that the cost of finding a solution (or proving none exists) 
depends essentially on the parameter 
\begin{equation}
\beta=\xi/\mu \, ,
\end{equation}
which characterizes the average number of constraints 
{\em per variable}~\cite{bib_Cheeseman91,bib_Kirkpatrick94,bib_Williams94}.
Specifically, the problem solving difficulty exhibits
a ubiquitous easy-hard-easy pattern, 
with the most difficult problem instances clustered
around a critical value of $\beta$ given, approximately, by
\begin{equation}
\beta_c = b^k \log(b)
\end{equation}
assuming $b^k\gg 1$ for simplicity.
This phenomenon, akin to a {\em phase transition} in physical
systems~\cite{bib_Kirkpatrick94,bib_Williams94}, persists across many
different sophisticated algorithms. The average case
complexity for a fixed $\beta$ is therefore believed to be a more informative
measure of computational complexity than either worst case
or average case complexity.\footnote{The motivation for investigating
the complexity of NP-complete problems in term of $\beta$ is that
worst case analyses can be misleading because they tend to focus on
atypical problem instances. Similarly, average case analyses can be
misleading because they are sensitive to the choice of the ensemble of
problem instances over which the average is computed. Such an ensemble
may contain for example an exceedingly large number of easy instances.}
It is the measure that we will use in the rest of this paper
for estimating the scaling of the complexity
of our improved quantum search algorithm (as well as the 
corresponding classical search algorithm).
\par

\subsection{Average computational complexity of a classical algorithm}

Let us describe a simple classical algorithm for a tree search problem
that exploits the structure of the problem by use of nesting. As pictured
in Fig.~\ref{fig_tree},
the key idea is to perform a preliminary search 
through a space of {\em partial}
solutions in order to avoid a search through the entire space
at the bottom of the tree. By definition,
a partial solution at level $i$ in the tree assigns values to
a subset of $i$ so-called {\em primary} variables ($x_1,\cdots,x_i$), 
which we denote as $A$.
The subset of {\em secondary} variables ($x_{i+1},\cdots,x_\mu$),
denoted as $B$, corresponds to 
the variables to which we assign a value only when extending
the partial solutions (i.e., when considering the descendants
of the partial solutions). 
In general, any partial solution can be tested against a part
of the constraints, namely just those constraints involving
the primary variables $A$. 
A partial solution that satisfies all these (testable) constraints
can be viewed as a {\em could-be} solution in the sense 
that all solutions at the bottom of the tree (at level $\mu$)
must be descendants of could-be's.
A classical search can be speeded up 
by terminating search along paths that are {\em not} descendants
of a could-be, thereby avoiding to search through the entire
space. The following algorithm can be used:
\begin{itemize}
\item Find a could-be solution at level $i$ in the tree. For this
purpose, choose repeatedly a random partial solution at level $i$,
until it satisfies the testable constraints. 
\item For each could-be solution, check exhaustively (or by use of
a random search) all its descendants at the bottom of the tree
(level $\mu$) for the presence of a possible solution.
\end{itemize}
This is clearly not a sophisticated algorithm. It
amounts to nesting the search for a successful descendant at level $\mu$
into the search for a could-be at level $i$. Nevertheless, it does
exploit the problem structure by using the knowledge gleaned from
the search at level $i$ to focus the search at level $\mu$. 
By finding a quantum analog of this algorithm 
(cf. Sec.~\ref{sect_nested_quant}), we
will, therefore, be able to address the impact of problem structure
on quantum search.
\par

Let us estimate the expected cost of running this algorithm. 
This cost consists essentially of three components: $n$, the cost
of finding a consistent partial solution (a could-be) at level $i$
in the tree, $m$, the cost of the subsequent search among its
descendants at level $\mu$, and $r$, the number of repetitions
of this whole procedure before finding a solution.
The search space for partial solutions (assignment of the primary 
variables $A$) is of size $d_A=b^i$. Assuming that 
there are $n_A$ could-be solutions at level $i$ (i.e.,
partial solutions that can lead to a solution),
the probability of finding one of them
by using a random search is thus $n_A/d_A$. 
Thus, one needs of the order of 
\begin{equation}
n = d_A/n_A
\end{equation}
iterations to find one could-be solution.
The descendants of a could-be are obtained by assigning a value
to the subset of {\em secondary} variables $B$, of size $d_B=b^{\mu-i}$
(each could-be has $d_B$ descendants).
Thus, searching through the entire
space of descendants of a could-be requires on average
\begin{equation}
m=d_B
\end{equation}
iterations.
If the problem admits a single solution,
this whole procedure needs to be repeated $n_A$ times, on average, 
since we have $n_A$ could-be solutions. More generally, if 
the number of solutions of the problem is given by $n_{AB}$,
this procedure must only be repeated
\begin{equation}
r=n_A/n_{AB}
\end{equation} 
times in order to find a solution with a probability
of order 1. Thus, the total number of iterations
required to find a solution of an average instance 
is approximately equal to
\begin{equation}  \label{eq_Tc}
T_c \simeq r(n+m) = {d_A+n_A d_B \over n_{AB}}
\end{equation}
This corresponds to an improvement over a naive {\em unstructured} 
search algorithm.
Indeed, the cost of a naive algorithm that does not exploit structure
is simply $d_{AB}/n_{AB}$, where
$d_{AB}=d_A d_B = b^\mu$ is the dimension of the total search space.
\par

The first term in the numerator of Eq.~(\ref{eq_Tc})
corresponds to the search for could-be solutions
in a space of partial solutions of size $d_A$ 
(shaded area at level $i$ in Fig.~\ref{fig_tree}),
while the second term corresponds to
the search for actual solutions among all the descendants of
the $n_A$ could-be solutions, each of them
having $d_B$ descendants (shaded area at level $\mu$
in Fig.~\ref{fig_tree}). The denominator 
in Eq.~(\ref{eq_Tc}) accounts for a problem admitting more than one solution.
We will see in Sec.~\ref{sect_nested_quant}
that the quantum counterpart
of Eq.~(\ref{eq_Tc}) involves taking the square root 
of $n$, $m$, and $r$, which essentially results in quantum-mechanical
square-root speedup over this classical algorithm.
\par

To make the estimate of this classical average-case complexity 
more quantitative, let $p(i)$ be the probability that
a partial solution at level $i$ is ``good'' (i.e., that it satisfies
all the testable constraints). In Appendix~\ref{app_p(i)},
we provide an asymptotic estimate of $p(i)$ for an average instance
of a large problem ($\mu\to\infty$) with a fixed value of the
parameter $\beta$. Recall that, if we want to preserve 
the difficulty while considering the limit of large problems,
$\beta$ must be kept constant. This is necessary for 
the complexity measure that we consider in this paper,
as mentioned before.
Thus, by making use of this asymptotic estimate of $p(i)$,
we can approximate the expected number of could-be solutions
at the $i$th level, $n_A=p(i)b^i$, and the expected number of
solution at the bottom of the tree, $n_{AB}=p(\mu)b^\mu$.
Therefore, the average search time of
the classical algorithm to find the first solution
is approximately equal to
\begin{equation}   \label{eq_Tclass}
T_c(i) = { b^i + p(i) b^\mu \over p(\mu) b^\mu }
\end{equation}
This is essentially the cost for finding one solution 
(which basically requires checking all the partial
solutions for a could-be, and subsequently checking the descendants of
all these could-be solutions) divided by the expected number
of solutions.\footnote{The denominator of Eq.~(\ref{eq_Tclass})
is the number of solutions at the bottom of the tree. For a problem
of maximum difficulty ($\beta=\beta_c$), it is shown in 
Appendix~\ref{app_p(i)} that $p(\mu)=b^{-\mu}$, i.e., the problem 
admits a single solution on average.}
\par

Equation~(\ref{eq_Tclass}) yields an approximate cost measure
for our classical nested search algorithm as a function
of the level of the ``cut'', $i$. An important question now
is where to cut the search tree? If one cuts the tree too high
(searching for could-be solutions at small $i$), one is unlikely
to learn anything useful as most partial solutions 
will probably be ``goods'', allowing for little discrimination
between solutions and non-solutions.
In other words, the second term in the numerator of Eq.~(\ref{eq_Tclass})
dominates since $p(i)$ is close to 1, i.e.,
there are many could-be solutions at level $i$. Searching
for could-be solutions is thus fast (the space of primary variables
is of size $b^i$ only), but those partial solutions are of little use
for singling out the actual solutions. The cost of the search
among the descendants of those partial solutions
is then high. Conversely, if one cuts the tree too deep 
(searching for could-be solutions at large $i$), although this would enhance
discrimination between solutions and non-solutions, the search space
for the primary variables would be almost as large as the entire
space. Then, the first term dominates the scaling as the search 
for could-be solutions becomes time-consuming.
It is therefore apparent that, for a typical problem instance,
there ought to be an optimal level at which to cut.
\par

We can estimate the optimal level by finding the value of $i$
that minimizes the classical 
computation time $T_c(i)$ for a given value of $\mu$, $b$, $\xi$, and $k$,
using the functional form for $p(i)$. This is done in 
Appendix~\ref{app_classcomplexity}, where we estimate the {\em asymptotic}
behavior of the location of the optimal cut level as a function of $\beta$,
as larger problems are considered ($\mu\to\infty$).
We then calculate the corresponding asymptotic scaling of $T_c$
for large problems. The result is that
the computation cost of running the classical nested search algorithm
scales as 
\begin{equation}
T_c \simeq b^{\alpha\mu} 
\end{equation}
for a search space
of dimension $d=b^\mu$, where $\alpha<1$ is a constant depending
on the problem considered. (More generally, this constant also depends
on the number of nesting levels, but we have considered a single level
of nesting in this Section.)
As we will see, the structured quantum search algorithm
that we present in Sect.~\ref{sect_nested_quant}
has a computational cost
of order of $\sqrt{b^{\alpha\mu}}$, in agreement
with the idea that a square root speedup is the best that can be
achieved by quantum mechanics. The focus of this paper,
however, is to show explicitly {\em how} a quantum algorithm
can be implemented that reaches this
maximum speedup over the classical algorithm discussed above. 
Interestingly enough, the quantum complexity 
of our nested algorithm scales
then as a power of the dimension of the search space $d=b^\mu$
that is less than 1/2. Structured quantum search therefore offers
a significant speedup over both structured classical search
and unstructured quantum search.
\par

\section{Unstructured quantum search}
\label{sect_unstruc_qu}

Let us first review the standard {\em unstructured} quantum search
algorithm~\cite{bib_grover0,bib_grover1}.
Consider a Hilbert space of dimension $d$ in which each basis $|x\rangle$
state ($x=1,\cdots d$) corresponds to a candidate solution of a
search problem. Any search problem can be recast as the problem
of finding the value(s) of $x$ at which an ``oracle'' function $f(x)$
is equal to one (this function being zero elsewhere).
We start the quantum search process from an arbitrary basis state
$|s\rangle$, and the goal is to reach a solution
(or target) basis state $|t\rangle$, with $f(t)=1$,
in a shortest computation time.
More precisely, if there is a single solution (or target state),
the goal is to reach a state that has an amplitude of order~1 in $|t\rangle$,
so that a measurement of this state gives the solution
with probability of order~1. (If there are $r$ solutions, the goal
is to reach a superposition of the states $|t\rangle$, each
with an amplitude of order $r^{-1/2}$.)
\par

The quantum search algorithm we discuss below is in fact an immediate
extension of the original one~\cite{bib_grover0,bib_grover1},
where an arbitrary unitary transformation is used instead
of the Walsh-Hadamard transformation~\cite{bib_grover2}.
Assume that we have at our disposal a quantum circuit that
performs a particular unitary operation $U$. If this operation
connects the starting state $|s\rangle$ to the target state $|t\rangle$,
i.e., $\langle t|U|s\rangle \ne 0$,
then this operation can be used classically to find the target. Indeed,
if we measure the system after applying $U$, the probability of obtaining
the solution $|t\rangle$ is obviously $|\langle t|U|s\rangle|^2$. Thus,
on average, we need to repeat this experiment $|\langle
t|U|s\rangle|^{-2}$ times to find the solution with probability
of order 1. We will show now that, using a quantum algorithm, 
it is possible to reach the target state
$|t\rangle$ in a number of steps of order $|\langle t|U|s\rangle|^{-1}$
only, which represents a huge speedup 
provided that $|\langle t|U|s\rangle|\ll 1$
(this corresponds to the situation of interest where the search space
is very large).
\par

The idea behind a quantum search algorithm is to {\em postpone} the
measurement, and keep a superposition of quantum states
throughout the algorithm. Only at the end, a measurement is performed.
Let us define the unitary operation
\begin{equation}  \label{eq_defQ}
Q = -U I_s U^{\dagger} I_t 
= - U \, e^{\displaystyle i\pi P_s}\, U^{\dagger} \, e^{\displaystyle i\pi P_t}
\end{equation}
where $P_s=|s\rangle\langle s|$
and $P_t=|t\rangle\langle t|$ are projection operators on
$|s\rangle$ and $|t\rangle$, respectively.
The two unitary operators $I_s=\openone-2P_s$ and $I_t=\openone-2P_t$
perform a controlled-phase operation:
applying $I_s$ (or $I_t$) on a state $|x\rangle$ flips its phase if
$x=s$ (or $x=t$), and leaves it unchanged otherwise.
Note that the target state $|t\rangle$ is of course not available
(it is what we are searching for).
Instead, we have at our disposal the quantum circuit (or ``oracle'') 
that computes the function $f(x)$, and we can use it to implement
the circuit for $I_t$. Thus, we have
$I_t|x\rangle = (-1)^{f(x)}|x\rangle$ for all state $|x\rangle$. The
circuit for $I_s$ does not require the function $f(x)$ and is trivial.
The principle at the heart of quantum search is to apply the operation $Q$
repeatedly in order to {\em amplify} the target component $|t\rangle$,
starting from $U|s\rangle$. 
This quantum {\it amplitude amplification}~\cite{bib_brassard}
can be understood by noting that, after applying $U$ to the starting
state $|s\rangle$, the repeated applications of $Q$
essentially rotate this state into the target state $|t\rangle$
at an angular velocity that is {\em linear} 
in the number of iterations. More specifically,
using $Q=-\openone+2|t\rangle\langle t|+2U|s\rangle\langle s|U^\dagger
-4U|s\rangle\langle s|U^\dagger|t\rangle\langle t|$,
we can see that $Q$ preserves the
two-dimensional subspace spanned by $U|s\rangle$ and $|t\rangle$,
namely
\begin{equation}
Q \left(
\begin{array}{c}
U|s\rangle \\
|t\rangle
\end{array}  \right) = \left(
\begin{array}{cc}
1-4\,|\langle t|U|s\rangle|^2 & 2\langle t|U|s\rangle \\
-2\langle t|U|s\rangle^*    & 1
\end{array}  \right) \; \left(
\begin{array}{c}
U|s\rangle \\
|t\rangle
\end{array}  \right)
\end{equation}
Therefore, in the case where $|\langle t|U|s\rangle| \ll 1$, the states
$U|s\rangle$ and $|t\rangle$ are almost orthogonal, and $Q$
tends to a rotation matrix of angle $2|\langle t|U|s\rangle|\ll 1$.
Indeed, keeping only the first-order terms in 
$u \equiv \langle t|U|s\rangle$, we obtain
\begin{eqnarray}
Q \left(
\begin{array}{c}
U|s\rangle \\
|t\rangle
\end{array}  \right) &\simeq& \left(
\begin{array}{cc}
1 & 2u \\
-2u^*  & 1
\end{array}  \right) \; \left(
\begin{array}{c}
U|s\rangle \\
|t\rangle
\end{array}  \right)  \nonumber \\
&\simeq& \exp\left(
\begin{array}{cc}
0 & 2u \\
-2u^*  & 0
\end{array}  \right) \; \left(
\begin{array}{c}
U|s\rangle \\
|t\rangle
\end{array}  \right)
\end{eqnarray}
We can then easily approximate the operation
of $Q^n$ in the subspace spanned by $U|s\rangle$ and $|t\rangle$:
\begin{eqnarray}  \label{eq_sinus}
Q^n  \left(
\begin{array}{c}
U|s\rangle \\
|t\rangle
\end{array}  \right) &\simeq& \exp\left(
\begin{array}{cc}
0 & 2nu \\
-2nu^* & 0 
\end{array}  \right) \; \left(
\begin{array}{c}
U|s\rangle \\
|t\rangle
\end{array} \right) \nonumber \\
&\simeq& \left(
\begin{array}{cc}
\cos (2n|u|) & {\displaystyle u\over\displaystyle |u|} \sin(2n|u|) \\
-{\displaystyle u^*\over\displaystyle |u|} \sin(2n|u|) & \cos(2n|u|)
\end{array}  \right) \; \left(
\begin{array}{c}
U|s\rangle \\
|t\rangle
\end{array}  \right)    \label{eq_9}
\end{eqnarray}
implying that the amplitude of the target state $|t\rangle$
after $n$ iterations is
\begin{equation}        \label{eq_10}
\langle t | Q^n U | s \rangle \; \simeq \; 
u \, \cos(2n|u|) + {\displaystyle u\over\displaystyle |u|} \sin(2n|u|)
\end{equation}
These last expressions are only asymptotically valid,
at the limit of small $|u|$. The exact expressions 
for Eqs.~(\ref{eq_9}) and (\ref{eq_10}) in terms 
of Chebyshev polynomials can be found in Appendix~\ref{app_chebyshev}.
\par

Consider first the case of a small rotation angle. From Eq.~(\ref{eq_10}), 
we see that if we iterate the application of $Q$ on $U|s\rangle$,
the amplitude of $|t\rangle$ grows {\em linearly} 
with the number of iterations $n$ provided 
that the total angle $2n|u| \ll 1$:
\begin{equation}
\langle t | Q^n U | s \rangle \; \simeq \; 
(1+2n) \; \langle t|U|s\rangle
\end{equation}
Consequently, if we measure the system after $n$ iterations,
the probability $p(n)$ of finding the solution grows {\em quadratically}
with $n$, as $p(n) \sim n^2 |\langle t|U|s\rangle|^2$. 
This is a great improvement compared
to the linear scaling of the classical algorithm
consisting in repeating $n$ times the measurement of $U|s\rangle$, namely 
$p(n) \sim n |\langle t|U|s\rangle|^2$. This is the quadratic
amplification effect provided by quantum mechanics.
\par

Now, consider the goal of reaching the target state $|t\rangle$
using this operator $Q$. From Eq.~(\ref{eq_sinus}) we see that,
starting from the state $U|s\rangle$, we need to apply $Q$
until we have rotated it by an angle of about $\pi/2$
in order to reach $|t\rangle$. At this time only, one
measures the system and gets the desired solution with a probability 
of order 1. The number of iterations required to rotate
$U|s\rangle$ into the solution $|t\rangle$ is thus
\begin{equation}    \label{eq_quant-r}
n \simeq {\pi \over 4} \; |\langle t|U|s\rangle|^{-1}
\end{equation}
and scales as the {\em square root} of the classical time.
It is worth noting that the amplitude of any state $|x\rangle$ orthogonal
to the target $|t\rangle$ is given by
\begin{equation}
\langle x | Q^n U | s \rangle \simeq
\cos(2n|u|) \, \langle x|U|s \rangle 
\end{equation}
so that
$\langle x | Q^n U | s \rangle \simeq \langle x|U|s \rangle$
for small angles. Thus, the amplitude of non-solutions is {\em not} amplified
by applying $Q$ repeatedly, so that the quantum search algorithm
selectively amplifies the solutions only.
\par

Thus, we have described here a general technique for achieving
a quantum-mechanical square-root speedup of a search algorithm
relying on {\em any} unitary transformation $U$~\cite{bib_grover2}.
The quantum search algorithm can be simply viewed as
a rotation from $U|s\rangle$ to $|t\rangle$ based on the
repeated operation of $Q$, followed by a measurement.
In the above discussion, the search operator $U$ can be 
arbitrary, {\em provided} it connects $|s\rangle$ and $|t\rangle$.
In the case of an {\em unstructured} search problem, 
as we have no {\it a~priori} knowledge about where the solution
is located, the most natural choice for $U$ is the Walsh-Hadamard
transformation $H$~\cite{bib_grover0,bib_grover1}:
\begin{equation}
H|x\rangle = {1\over\sqrt{d}} \sum_{y=0}^{d-1}  
(-1)^{\overline{x}\cdot\overline{y}} |y\rangle
\end{equation}
where $\overline{x}\cdot\overline{y}=\sum_{i=0}^{d-1} x_i y_i ({\rm mod}~2)$,
with $x_i$ ($y_i$) being the binary digits of $x$ ($y$).\footnote{Here 
and below, we assume that $d$ is a power of 2 for simplicity.}
Indeed, $U=H$ does not bias the search towards a particular 
candidate solution since 
$H|s\rangle$ has the same (squared) amplitude in all the candidate solutions,
so that the search starts from a uniform distribution of all states.
Applying $U=H$ to an arbitrary state $|s\rangle$, we see that 
\begin{equation}
\langle t|H|s\rangle = \pm 1/\sqrt{d}
\end{equation}
for all possible target state $|t\rangle$. 
Thus, according to Eq.~(\ref{eq_quant-r}),
the number of iterations in the quantum search algorithm relying on $H$ 
is $O(\sqrt{d})$~\cite{bib_grover0,bib_grover1},
whereas a classical search algorithm obviously requires $O(d)$ steps. 
When there are multiple target states (the
problem admits several solutions), it can be shown that
the quantum computation time becomes $O(\sqrt{d/r})$, where $r$ is the
number of solutions~\cite{bib_boyer}. The classical counterpart
is then simply $O(d/r)$.
\par

For a {\em structured} search problem, however, it is natural
to use the knowledge of the structure in order to choose
a better $U$. Indeed, if we have partial knowledge about
where the solutions are,
we can exploit it to {\em bias} the search in such a way
that $U|s\rangle$ has larger amplitudes in states which are more
probable to be solutions. This is the focus of the present paper.
It has been shown recently that an arbitrary (non-uniform) initial amplitude
distribution can be used as well with the standard quantum search
algorithm, resulting in a $O(\sqrt{d/r})$ quantum 
computation time~\cite{bib_biron}. This seems to indicate that
the scaling remains in $O(\sqrt{d})$ even if we use our knowledge
about the problem by biasing the initial distribution.
In contrast, we will show in Sec.~\ref{sect_nested_quant} that the use of 
a {\em nested} quantum search algorithm can result in a power law in $d$
with an exponent that is {\em smaller} than 1/2. 
The key idea is that $U$ is not fixed {\it a priori},
but is rather obtained ``dynamically'' by the quantum algorithm itself, 
depending on the particular instance.
In short, the standard search algorithm is used
to {\em construct} an effective search operator $U$ (or a non-uniform initial
distribution) which, itself, is nested within another quantum
search algorithm. In other words, we apply quantum search ``recursively'':
the operator $(-H I_s H I_t)^n H$ resulting from the nested search algorithm
based on $H$ is used as a better search operator $U$ 
for a quantum search at an upper level of hierarchy.

\section{Nested quantum search on structured problems}
\label{sect_nested_quant}

\subsection{The core quantum algorithm}
\label{sect_core}

\begin{figure}
\caption{Schematic representation of stages (i) and (ii) 
of the quantum algorithm. These stages partially amplify the solution
states, and can be nested into a standard quantum search algorithm (iii)
in order to speedup the amplification of the solutions.}
\vskip 0.25cm
\centerline{\psfig{file=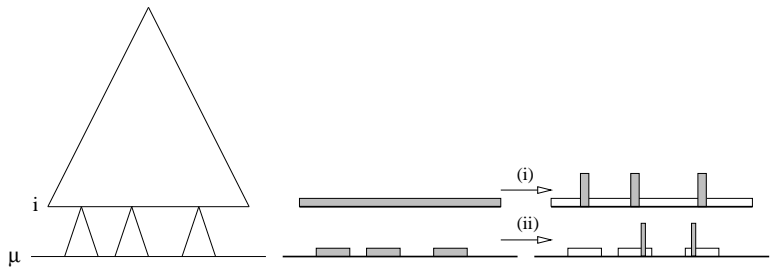,width=4.0in,angle=-90}}
\label{fig_idea}
\vskip -0.25cm
\end{figure}

Assume that the Hilbert space of our search problem
is the tensor product of two Hilbert spaces ${\cal H}_A$ and 
${\cal H}_B$. As before, $A$ denotes the set of primary variables, that
is, the variables to which we assign a value in the first stage.
The partial solutions correspond to definite values for these
variables. Thus, ${\cal H}_A$ represents the search space 
for partial solutions (of dimension $d_A$). The set of secondary 
variables, characterizing the extensions of partial solutions,
is denoted by $B$, and the corresponding Hilbert space
${\cal H}_B$ is of dimension $d_B$. Let us briefly describe the
quantum algorithm with a single nesting level (the counterpart of the
classical algorithm of Sect.~\ref{sect_nested_class}):
\begin{itemize}
\item The first stage (i) consists in constructing
a superposition (with equal amplitudes) of all the could-be solutions
at level $i$ by use of the standard unstructured search algorithm
based on $H$. 
\item Then (ii), one performs a subsequent quantum search 
in the subspace of the descendants of {\em all} the could-be partial
solutions, simultaneously. This second stage is achieved 
by using the standard quantum search algorithm with, as an input,
the {\em superposition} of could-be solutions resulting from
the first stage. The overall yield of stages (i) and (ii) 
is a superposition of all states where
the solutions have been partially amplified with respect to
non-solutions.
\item The final procedure (iii) consists
of nesting stages (i) and (ii)--- using them as a search 
operator $U$---inside a higher-level quantum search algorithm
until the solutions get maximally amplified, at which point
a measurement is performed. This is summarized in Fig.~\ref{fig_idea}.

\end{itemize}
\par

Let us now follow in more details the evolution of the quantum state
by applying this quantum nested algorithm, and estimate the number
of iterations required. The starting state of the search is denoted
as $|s,s'\rangle$, where $|s\rangle$ (lying in ${\cal H}_A$)
and $|s'\rangle$ (lying in ${\cal H}_B$) are just the initial state
of two different
parts of the same, single, quantum register which is large enough to
hold all the potential solutions in the total search space (i.e.. all
the $b^\mu$ leaf nodes of the search tree at level $\mu$).
Register $A$ stores the starting state at 
an intermediate level $i$ in the tree, while
register $B$ stores the continuation of that state 
at level $\mu$. In other words, $A$ holds partial solutions
and $B$ their elaboration in the leaves of the tree.
\par
\bigskip

(i) The first stage of the algorithm consist in a standard quantum
search for {\em could-be} partial solutions $|c\rangle$ 
at level~$i$, that is, states in subspace ${\cal H}_A$
that do not violate any (testable) constraint. 
We start from state $|s\rangle$ in subspace ${\cal H}_A$, 
and apply a quantum search based on the Walsh-Hadamard transformation $H$
since we do not have {\it a priori} knowledge about the location
of could-be solutions. Using
\begin{equation}
\langle c | H | s \rangle = \pm 1 / \sqrt{d_A}
\end{equation}
we can perform an amplification of the components $|c\rangle$
based on $Q = -H I_s H I_c$ where
\begin{eqnarray}
I_s&=&\exp(i\pi |s\rangle\langle s| ) \label{eq_definI_s} \\
I_c&=&\exp(i\pi \sum_{c\in C} |c\rangle\langle c|)
\end{eqnarray}
The states $|c\rangle$ correspond to the could-be partial solutions
in ${\cal H}_A$ (assignment of the primary variables that could lead
to a solution), and belong to the subset $C=\{c_1,\cdots,c_{n_A}\}$.
We assume that there are $n_A$ could-be partial solutions,
with $1\ll n_A\ll d_A$.
The quadratic amplification of these could-be solutions, starting
from $|s\rangle$, is reflected by
\begin{equation}
\langle c| Q^n H| s\rangle 
\simeq n \; \langle c| H| s\rangle 
\simeq n / \sqrt{d_A}
\end{equation}
for small rotation angle. Thus, applying $Q$ sequentially, we can construct
a superposition of all the could-be solutions $|c\rangle$,
each with an amplitude of order $\sim 1/\sqrt{n_A}$. 
The required number of iterations of $Q$ scales as
\begin{equation}
n \simeq \sqrt{d_A / n_A}
\end{equation}
This amplitude amplification process can equivalently be
described in the joint Hilbert space ${\cal H}_A\otimes{\cal H}_B$,
starting from the product state $|s,s'\rangle$,
where $|s'\rangle$ denotes an arbitrary starting state in ${\cal H}_B$,
and applying $(Q\otimes \openone)$ sequentially:
\begin{equation}   \label{eq_1ststep}
\langle c,s'| (Q\otimes \openone)^n (H\otimes \openone) | s,s'\rangle 
= \langle c|Q^n H| s\rangle
\sim n / \sqrt{d_A}
\end{equation}
Here and below, we use the convention that the left (right) term in a tensor
product refers to subspace $A$ ($B$).
\par


\bigskip

(ii) The second stage of the algorithm is a standard quantum search 
for the secondary variables $B$ in the subspace of the ``descendants''
of the could-be solutions that have been singled out in stage (i).
As before, we can use the search operator
$H$ that connects extended could-be solutions $|c,s'\rangle$
to the actual solutions or target states $|t,t'\rangle$ in the joint
Hilbert space:
\begin{equation}
\langle t,t' | (\openone\otimes H) | c,s' \rangle = 
\langle t|c\rangle \, \langle t'|H|s'\rangle
= \pm \delta_{\displaystyle c,t} / \sqrt{d_B}
\end{equation}
Note that, this matrix element is non-vanishing only for
could-be states $|c\rangle$ that lead to an actual solution.
Define the operator $R = -(\openone\otimes H I_{s'} H) I_t$, with
\begin{eqnarray}
I_{s'}&=& \exp\big( i\pi 
      |s'\rangle\langle s'| \big)   \label{eq_definI_s'}\\
I_{t\phantom{'}}&=&\exp\big( i\pi \sum_{(t,t')\in T}
      |t,t'\rangle\langle t,t'| \big)              \label{eq_defin_I_t}
\end{eqnarray}
where $T$ is the set of solutions $|t,t'\rangle$ 
at the bottom of the tree, and
$\#(T)=n_{AB}$, i.e., the problem admits $n_{AB}$ solutions. 
We can apply the operator $R$ sequentially 
in order to amplify a target state $|t,t'\rangle$, namely
\begin{equation}   \label{eq_2ndstep_pre}
\langle t,t' | R^m (\openone\otimes H) | c,s' \rangle 
\simeq
\left\{ \begin{array}{l @{\qquad} l}
m \; \langle t,t' | (\openone\otimes H) | c,s' \rangle
 & {\rm if~}c=t\\
\langle t,t' | (\openone\otimes H) |c,s'\rangle
 & {\rm if~}c\ne t
\end{array} \right.
\end{equation}
for small rotation angle. Note that, for a could-be state $|c\rangle$ that
does not lead to a solution ($c\ne t$), we have 
$I_t|c,x\rangle=|c,x\rangle$ for all $x$, so that
$R^m (\openone\otimes H) | c,s' \rangle 
= (-\openone\otimes H I_{s'} H)^m (\openone\otimes H)|c,s'\rangle
= (\openone\otimes H)|c,s'\rangle$,
and the matrix element is not amplified by $m$
compared to the case $c=t$. In other words,
no amplification occurs in the space of descendants of could-be partial
solutions that do not lead to an actual solution. 
Thus, Eq.~(\ref{eq_2ndstep_pre}) results in
\begin{equation}   \label{eq_2ndstep}
\langle t,t' | R^m (\openone\otimes H) | c,s' \rangle 
\simeq {m \over \sqrt{d_B}} \delta_{\displaystyle c,t} 
\end{equation}
Assuming that, among the descendants of each could-be
solution $|c,s'\rangle$,
there is either zero or one solution, we need to iterate
$R$ of the order of 
\begin{equation}
m \simeq \sqrt{d_B}
\end{equation}
times in order to maximally amplify each solution.
We then obtain
a superposition of the solution states $|t,t'\rangle$, each with an amplitude
$\sim 1/\sqrt{n_A}$. This can also be seen by
combining Eqs.~(\ref{eq_1ststep}) and (\ref{eq_2ndstep}),
and using the resolution of identity
$\openone=\sum_{x,y}|x,y\rangle\langle x,y|$:
\begin{eqnarray}
\langle t,t' | \underbrace {R^m (\openone\otimes H) 
               (Q\otimes\openone)^n (H\otimes\openone) }_U |s,s'\rangle
&=&\sum_{x,y} \langle t,t' | R^m (\openone\otimes H) |x,y\rangle \,
\langle x,y |(Q\otimes\openone)^n (H\otimes\openone) |s,s'\rangle \nonumber\\
&=&\langle t,t' | R^m (\openone\otimes H) |t,s'\rangle \,
\langle t,s' |(Q\otimes\openone)^n (H\otimes\openone) |s,s'\rangle \nonumber\\
&\simeq&  (m /\sqrt{d_B}) (n / \sqrt{d_A}) \nonumber\\
&\simeq&  1/\sqrt{n_A}
\end{eqnarray}
Thus, applying the operator $Q^n$ followed by the operator $R^m$
connects the starting state  $|s,s'\rangle$ to each of the solutions
$|t,t'\rangle$ of the problem with a matrix element of order
$\sim  1/\sqrt{n_A}$.
\par

\bigskip

(iii) The third stage consists in using
the operator $U\equiv R^m (\openone\otimes H) 
                     (Q\otimes\openone)^n (H\otimes\openone)$
resulting from steps (i) and~(ii) as a search operator for
a higher-level quantum search algorithm, in order to further
amplify the superposition of $n_{AB}$ target (or solution)
states $|t,t'\rangle$.
The goal is thus to construct such a superposition where each solution
has an amplitude of order $\sim 1/\sqrt{n_{AB}}$. As before,
we can make use of the operator 
$S=-U (I_s\otimes I_{s'}) U^{\dagger} I_t$ where
$I_s$, $I_{s'}$, and $I_t$ are defined in Eqs.~(\ref{eq_definI_s}), 
(\ref{eq_definI_s'}), and (\ref{eq_defin_I_t}),
in order to perform amplification according to the relation
\begin{equation}
\langle t,t'| S^r U |s,s'\rangle
\simeq r \; \langle t,t'|U|s,s'\rangle
\simeq r/\sqrt{n_A}
\end{equation}
for small rotation angle. The number of iterations of $S$
required to maximally amplify the solutions is thus of the order of
\begin{equation}
r \simeq \sqrt{n_A\over n_{AB}}
\end{equation}
This completes the algorithm. At this point, it is sufficient to perform
a measurement of the amplified superposition of solutions. This yields
one solution $|t,t'\rangle$ with a probability of order 1.
\par

\begin{figure}
\caption{Circuit implementing the nested quantum search algorithm
(with a single level of nesting).
The upper set of quantum variables, 
initially in state $|s\rangle$, corresponds to
the primary variables $A$. The lower set of quantum variables, initially
in states $|s'\rangle$, is associated with the secondary variables $B$.
The quantum circuit makes use of controlled-phase 
gates $I_s=\exp(i\pi|s\rangle\langle s|)$,
$I_{s'}=\exp(i\pi|s'\rangle\langle s'|)$,
$I_c=\exp(i\pi\sum_{c\in C}|c\rangle\langle c|)$, and
$I_t=\exp(i\pi\sum_{(t,t')\in T}|t,t'\rangle\langle t,t'|)$,
and Walsh-Hadamard gates $H$. The entire operation of $U$ (exhibited
inside the dashed box) is repeated $r$ times. Note that 
$U^{-1}=U^{\dagger}$
corresponds to same the circuit as $U$ but read from right to left. }
\vskip 0.25cm
\centerline{\psfig{file=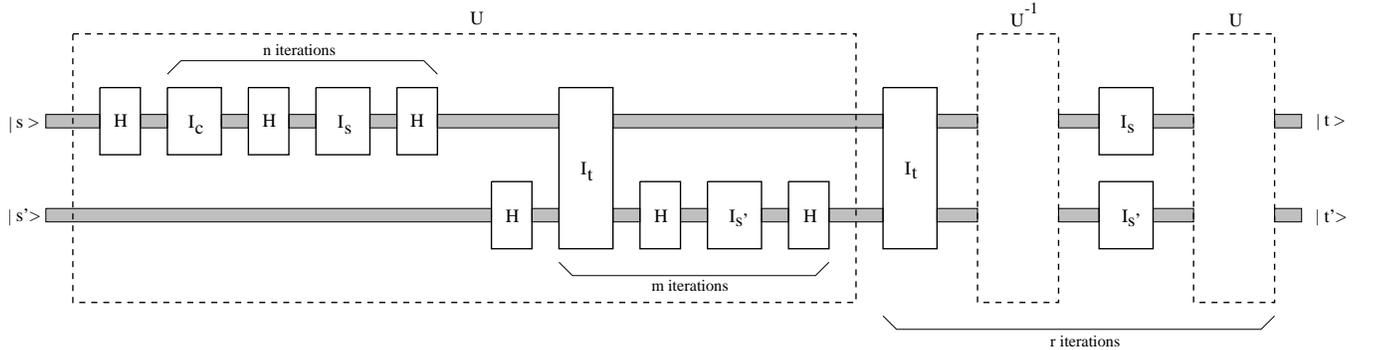,width=7.0in,angle=-90}}
\label{fig_algo}
\vskip -0.25cm
\end{figure}

In Fig.~\ref{fig_algo}, the quantum network that implements
this nested quantum search algorithm is illustrated.
Clearly, a sequence of two quantum search circuits (a search in the
$A$ space followed by a search in the $B$ space) is {\em nested} into
a global search circuit in the whole Hilbert space ${\cal H}_{AB}$.
This can be interpreted as a ``dynamical'' choice of
the search operator $U$ that is used in the global quantum search. 
This quantum nesting
is distinct from a procedure where one would try to
choose an optimum $U$ before running the quantum search by making use of the
structure {\em classically} (making several classical queries 
to the oracle) in order to speedup the resulting quantum search. 
Here, no measurement
is involved and structure is used at the quantum level.
\par

\subsection{Quantum average-case complexity}
\label{sect_qacc}

Let us estimate the total number of iterations, or more precisely the
number of times that a controlled-phase operator ($I_t$, which flips 
the phase of a solution, or $I_c$, which flips the phase
of a could-be partial solution) is used.
Since we need to repeat $r$ times the operation $S$, which itself
requires applying $n$ times $Q$ and $m$ times $R$, we obtain
for the quantum computation time
\begin{equation}
T_q\simeq r(n+m) \simeq { \sqrt{d_A}+\sqrt{n_A d_B} \over \sqrt{n_{AB}} } 
\end{equation}
This expression is the quantum counterpart of Eq.~(\ref{eq_Tc}),
and has the following interpretation.
The first term in the numerator corresponds to a quantum search
for the could-be partial solutions in space of size $d_A$. The second
term is associated with a quantum search of actual solutions
in the space of all the descendants of the $n_A$ could-be
solutions (each of them has a subspace of descendants of size $d_B$).
The denominator accounts for the fact that the total number of
iterations decreases with the square root of the number of solutions
of the problem $n_{AB}$, as in the standard quantum search
algorithm.
\par

Let us now estimate the scaling of the computation time
required by this quantum nested algorithm
for a large search space ($\mu\to\infty$). 
Remember that $\mu$ is the number of variables 
(number of nodes for the graph coloring problem) and $b$ 
is the number of values (colors) per variable.
As before, if we ``cut'' the tree
at level $i$ (i.e., assigning a value
to $i$ variables out of $\mu$ defines a partial solution), 
we have $d_A=b^i$ and $d_B=b^{\mu-i}$.  Also, we have 
$n_A=p(i)b^i$, and $n_{AB}=p(\mu)b^\mu$, where 
$p(i)$ is the probability of having a partial solution at level $i$ 
that is ``good'' in a tree of height $\mu$.
(The quantity $p(\mu)$
is thus the probability of having a solution in the total search space.)
We can reexpress the computation time as a function of $i$,
\begin{equation}  \label{eq_T(i)}
T_q(i)= {\sqrt{b^i} + \sqrt{p(i) b^\mu} \over \sqrt{p(\mu) b^\mu } }
\end{equation}
which is the quantum counterpart of Eq.~(\ref{eq_Tclass}).
In order to determine the scaling of $T_q$, we use the 
asymptotic estimate of $p(i)$ that is derived in Appendix~\ref{app_p(i)},
namely
\begin{equation}   \label{eq_p(i)}
p(i)= b ^ {\displaystyle  - \mu \left({\beta\over\beta_c}\right) 
                         \left({i\over \mu}\right)^k }
\end{equation}
Eq.~(\ref{eq_p(i)}) is a good approximation of $p(i)$ in the asymptotic regime,
i.e., when the dimension of the problem $\mu$ (or the number of variables)
tends to infinity. 
Remember that, in order keep the difficulty constant
when increasing the size of the problem,
we need to choose the number of constraints 
$\xi = \beta \mu$ when $\mu\to\infty$.\footnote{For the 
graph coloring problem, since $\xi = e b$ 
(where $e$ being the number of edges and $b$ the number of colors),
it implies that the number of edges must grow linearly with
the number of nodes for a fixed number of colors in order to
preserve the difficulty. In other words,
the average connectivity must remain constant.}
The constant $\beta$ corresponds to the average number of constraints 
{\em per variable}, and is a measure of the difficulty of the problem.
The difficulty is maximum when $\beta$ is close to
a {\em critical} value $\beta_c= b^k \log(b)$, where
$k$ is the size of the constraint (i.e., number of variables
involved in a constraint).
Note that $p(\mu)=b^{-\mu(\beta/\beta_c)}$, implying that the number
of solutions at the bottom of the tree is $n(\mu)=b^{\mu(1-\beta/\beta_c)}$.
Thus, if $\beta\simeq \beta_c$, we have $p(\mu)\simeq b^{-\mu}$,
so that the problem admits of the order of $n(\mu)\simeq 1$ solutions.
This corresponds indeed to the hardest case, where one is searching
for a single solution in the entire search space. When
$\beta<\beta_c$, however, there are less constraints and the problem
admits more than one solution, on average. If $\beta>\beta_c$,
the problem is overconstrained, and it typically becomes easier
to check the nonexistence of a solution.
\par

Now, plugging Eq.~(\ref{eq_p(i)}) into Eq.~(\ref{eq_T(i)}),
we obtain for the quantum computation time
\begin{equation}   \label{eq_T_q(i)}
T_q(i) 
\simeq {\sqrt{b^{\displaystyle i}} + 
        \sqrt{b^{\displaystyle \mu-\mu (\beta/\beta_c)(i/\mu)^k}}
  \over \sqrt{b^{\displaystyle \mu-\mu(\beta/\beta_c)}}  }
\end{equation}
Defining the {\em reduced} level on the tree as $x=i/\mu$, i.e.,
the fraction of the height of the tree at which we exploit the structure
of the problem, we have
\begin{equation}  \label{eq_T(x)}
T_q(x) = { a^{\displaystyle x} + a^{\displaystyle 1-(\beta/\beta_c)x^k}
\over a^{\displaystyle 1-\beta/\beta_c}  }
\end{equation}
where $a\equiv \sqrt{b^\mu}$.
Now, we want to find the value of $x$ that
minimizes the computation time $T_q(x)$, so we have to solve
\begin{equation}
(\beta/\beta_c)\, k x^{k-1} = a^{(\beta/\beta_c)x^k+x-1}
\end{equation}
For large $\mu$ (or large $a$), this equation asymptotically reduces to
\begin{equation}   \label{eq_equationforx}
(\beta/\beta_c)\, x^k + x - 1 = 0 
\end{equation}
The solution $x$ (with $0\le x\le 1$)
corresponds therefore to the reduced level
for which $T_q(x)$ grows asymptotically ($\mu\to\infty$)
with the smallest power in $b$.
Note that this optimum $x$ is such that both terms in the numerator
of Eq.~(\ref{eq_T_q(i)}) grow with the same power in $b$ (for large
$\mu$). This reflects that there is a particular fraction $x$ 
of the height of the tree where it is optimal to ``cut'', i.e., 
to look at partial solutions. 
The optimum computation time can then be written as
\begin{equation}  \label{eq_T(x)_solution}
T_q \simeq 
{2 a^{\displaystyle \alpha} \over  a^{\displaystyle 1-\beta/\beta_c}  }
\simeq {\sqrt{b^{\displaystyle \alpha \mu}} 
  \over \sqrt{b^{\displaystyle \mu(1-\beta/\beta_c)}}  }
\end{equation}
where the constant $\alpha<1$ is defined as the solution $x$ 
of Eq.~(\ref{eq_equationforx}).\footnote{We may ignore the prefactor 2
as it only yields an additive constant in the logarithm
of the computation time.}
Note that, for a search with several levels of
nesting, the constant $\alpha < x$, as we shall see
in Sect.~\ref{sect_multi}.
\par

Equation~(\ref{eq_T(x)_solution}) implies that
the scaling of the quantum search in a space of dimension $d=b^\mu$
is essentially $O(d^{\alpha/2})$
modulo the denominator (which simply accounts for the number of solutions).
In contrast, the standard {\em unstructured} quantum search algorithm
applied to this problem corresponds to $\alpha=x=1$, with
a computation time scaling as $T_q(\alpha=1)=O(d^{1/2})$.
This means that exploiting
the structure in the quantum algorithm results 
in a decrease of the power in $b$
by a coefficient $\alpha$: the power $1/2$ of the standard quantum search
is reduced to $\alpha/2$ for this nested quantum search algorithm.
Consider this result at $\beta=\beta_c$, i.e., when the difficulty of
the problem is maximum for a given size $\mu$. This is the most
interesting case since when $\beta<\beta_c$, 
the problem becomes easier to solve classically.
For $\beta=\beta_c$, the nested algorithm essentially scales as
\begin{equation}
T_q \simeq d^{\alpha/ 2} =\sqrt{b^{\alpha\mu}}
\end{equation}
where $\alpha=x<1$ with $x$ being the solution of $x^k + x - 1 = 0$, and
$d=b^\mu$ is the dimension of the search space.
This represents a significant improvement over the scaling of the
unstructured quantum search algorithm, $O(d^{1/2})$.
Nevertheless, it must be emphasized that the speedup with respect
to the computation time $O(d^\alpha)$ of the 
classical nested algorithm presented in Section~\ref{sect_nested_class}
is exactly a square root (cf. Appendix~\ref{app_classcomplexity}).
This implies that this nested quantum search algorithm 
is the {\em optimum} quantum version 
of this particular classical non-deterministic algorithm.
\par

For the graph coloring problem ($k=2$), we must solve the linear equation
of second order $x^2 + x - 1 = 0$, whose solution is simply 
$x=(-1+\sqrt{5})/2 = 0.6180$. (When $k>2$, the solution for $x$
increases, and tends to 1 for large $k$.)
This means that the level on the tree where it is optimal to use the structure
is at about 62\% of the total height of the tree, i.e., when assigning values
to about 62\% of the $\mu$ variables. In this case, the computation
time of the nested algorithm scales as $O(d^{\, 0.31})$, which is clearly
an important computational gain compared to $O(d^{\, 0.5})$.
\par

Consider the regime where $\beta < \beta_c$, i.e., there are
fewer constraints and therefore more than one solution on average,
so that the problem becomes easier to solve. 
For a given $k$, the solution $x$ of
Eq.~(\ref{eq_equationforx}) increases when $\beta$ decreases,
and tends asymptotically to 1 for $\beta\to 0$.
This means that we recover the {\em unstructured} quantum search
algorithm in the limit where $\beta\to 0$.
The denominator in Eq.~(\ref{eq_T(x)_solution}) increases,
and it is easy to check that the computation time
\begin{equation}
T_q \simeq \sqrt{b^{\displaystyle \mu(\alpha-1+\beta/\beta_c)}}
\end{equation}
decreases when $\beta$ decreases. As expected, 
the computation time of the nested algorithm approaches 
$O(\sqrt{d^{\beta/\beta_c}})$ as $\beta$ tends to 0 (or $x\to 1$), that is,
it reduces to the time of the standard unstructured
quantum search algorithm at the limit $\beta\to 0$.

\subsection{Quantum search with several levels of nesting}
\label{sect_multi}

The quantum algorithm described in Sect.~\ref{sect_core}
relies on a single level of nesting. Indeed, the search at
the bottom of the tree (level $\mu$) is speeded up by making
use of a search at level $i$ which determines the partial
solutions which are ``good''. Only the candidate solutions
which are descendants of these partial solutions
are examined in the search at level $\mu$. It should be
realized that these ``good'' partial solutions
at level $i$ are selected, themselves, by a {\em naive} search: 
stage (i) indeed amounts to use the standard unstructured search
based on $H$. In the corresponding classical
nested algorithm, this amounts to select a random partial solution
at level $i$ and check whether it is good. 

\par
It is natural that both the classical and the quantum algorithms
could be improved further if the search for good partial solutions
at level $i$ itself was made faster by making use of the structure
of the upper part of the tree (by examining partial solutions at level $j$,
with $j<i$, and considering only the descendants of the ``good'' ones). 
This leads to the notion of a search with several
levels of nesting (i.e., a nesting depth larger than one).
\par

In order to analyze the scaling achieved by several levels of nesting,
let us consider a search at level $i$ which corresponds to the
$n$-th nesting level. We suppose that this search 
relies itself on a search at level $j$, where $j<i<\mu$, which
corresponds therefore to the $n+1$-th nesting level.
Let $i=x_n\mu$ and $j=x_{n+1}\mu$, where $x_n$ and $x_{n+1}$
denote the reduced level on the tree 
at the $n$-th and $n+1$-th nesting level, respectively.
Assume that the quantum computation cost at level $j$ is given by
\begin{equation}
t(j) \simeq { \sqrt{b^{\alpha_{n+1} \, j }} \over \sqrt {p(j) b^j} }
\end{equation}
where $\alpha_{n+1}$ is the scaling coefficient at the $n+1$-th level
of nesting (level $j$ in the tree).
Using the structure at level $j$, the quantum computation cost
at level $i$ can be written as
\begin{eqnarray}
t(i) &\simeq&  {  \sqrt{p(j)b^j} \left( t(j) + \sqrt{b^{i-j}} \right)
            \over \sqrt{p(i) b^i}  }   \nonumber \\
&=&  {  \sqrt{b^{\alpha_{n+1}\, j} } + \sqrt{p(j)b^i}   
            \over \sqrt{p(i) b^i}  } 
\end{eqnarray}
By optimizing $j$ so that $t(i)$ is minimum, as before,
we obtain $j=x_{n+1}\mu$, where $x_{n+1}$ is a solution of
\begin{equation}  \label{eq_rec1}
(\beta/\beta_c) \, x_{n+1}^k + \alpha_{n+1} \, x_{n+1} - x_n =0
\end{equation}
with $0\le x_{n+1} \le 1$.
Defining the scaling coefficient $\alpha_n$ by
\begin{equation}  \label{eq_rec2}
\alpha_n x_n= \alpha_{n+1} x_{n+1}
\end{equation}
we see that the corresponding computation cost at level $i$
is given by
\begin{equation}
t(i) \simeq { \sqrt{b^{\alpha_{n} \, i }} \over \sqrt {p(i) b^i} }
\end{equation}
Thus, to determine the cost of the global algorithm,
we need to solve the set of recurrence 
equations~(\ref{eq_rec1})-(\ref{eq_rec2}) for $n=0,1,\cdots,N-1$,
where $N$ is the nesting depth ($N=1$ corresponds to the algorithm described
in Sect.~\ref{sect_core}). The boundary conditions
are $x_0=1$ (the upper level is a search for solutions
at the bottom of the tree, i.e., at level $\mu$)
and $\alpha_N=1$ (the innermost search at the $N$-th level of nesting
is supposed to be a naive search). These two conditions, together
with the $2N$ recurrence relations, uniquely determine
the variables $(x_0,x_1,\cdots x_N)$ and
$(\alpha_0,\alpha_1,\cdots,\alpha_N)$. The overall scaling of the
quantum search algorithm is $O(\sqrt{b^{\alpha_0\mu}})$, i.e.,
it is governed by $\alpha_0$ (the constant
that was denoted as $\alpha$ in the previous Sections). Note that
this entire calculation is also valid for a classical nested search
with several levels of nesting, except for the square root. Thus,
the speedup of the multi-nested quantum search algorithm
remains a square root if compared with the corresponding
multi-nested classical search algorithm.
\par

We show in Table~\ref{table1}
the values of the $x_n$'s and $\alpha_n$'s for 
an average instance of maximum 
difficulty ($\beta=\beta_c$) of
the graph coloring problem ($k=2$).
The scaling coefficient $\alpha_0$ decreases
with an increasing nesting depth $N$, implying that the
speedup over an unstructured search improves
by adding further nesting levels.
It should be emphasized, however,
that the formalism used to estimate the scaling throughout
this paper cannot be used for a large nesting depth $N$.
Indeed, the derivation of $p(i)$ essentially neglects the correlations
between partial solutions at any level in the tree which arise
because of their sharing a same ancestor. Thus, our cost estimate
for the multi-nested algorithm is only valid provided that $N\ll \mu$
(the fact that $\alpha_0\to 0$ when $N\to \infty$ is meaningless).
\par

\section{Conclusion}

There is considerable interest in the possibility of using
quantum computers to speedup the solution of NP-complete
problems given the importance of these problems 
in complexity theory and their ubiquity amongst practical
computational applications.
This paper presents an attempt in this direction
by showing that nesting the standard quantum search algorithm 
results in a faster quantum algorithm for structured search problems
such as the constraint satisfaction problem
than heretofore known. The key innovation is to cast
the construction of solutions of the problem as a quantum search
through a tree of partial solutions, which narrows a subsequent
quantum search at the next level in the search tree.
The corresponding computation time
scales exponentially, but with a reduced coefficient
that depends on the number of nesting levels and on
the problem. The speedup that is achieved is a square
root over the computation time of a corresponding classical
nested search algorithm, which represents therefore the appropriate benchmark.
Nevertheless, it is an exponential improvement with respect
to the time needed to solve the problem by use of
the standard unstructured quantum search algorithm.

\acknowledgements

NJC is supported in part 
by the NSF under Grant Nos. PHY 94-12818 and PHY 94-20470,
and by a grant from DARPA/ARO through the QUIC Program
(\#DAAH04-96-1-3086). CPW is supported by the NASA/JPL Center for
Integrated Space Microsystems (grant 277-3R0U0-0) 
and NASA Advanced Concepts (grant 233-0NM71-0).
NJC is {\it Collaborateur scientifique}
of the Belgian National Fund for Scientific Research.

\bigskip
\bigskip
\bigskip

\begin{table}
\begin{tabular}{l| c c | c c | c c | c c}
N & $x_0$ & $\alpha_0$ & $x_1$ & $\alpha_1$ & $x_2$ & $\alpha_2$ &
$x_3$ & $\alpha_3$ \\
\hline
1& 1.000 & 0.618 & 0.618 & 1.000 &-&-&-&-\\
\hline
2& 1.000 & 0.484 & 0.718 & 0.674 & 0.484 & 1.000 &-&-\\
\hline
3& 1.000 & 0.416 & 0.764 & 0.545 & 0.590 & 0.706 & 0.416 & 1.000 \\
\end{tabular}
\caption{Reduced level $x_n$ on the tree and corresponding
scaling coefficient $\alpha_n$ at the $n$-th level of nesting
for the graph coloring problem ($k=2$) at $\beta=\beta_c$.
The variable $N$ denotes the nesting depth, and $\alpha_0$
governs the scaling of the overall quantum (or classical)
algorithm. }
\label{table1}
\end{table}

\appendix

\section{Asymptotic probability of a node in a search
tree to be good}
\label{app_p(i)}

Let us derive an approximate functional form for $p(i)$, 
the probability that a node at level $i$ in the search tree is ``good''. 
The derivation is complicated by
the fact that the same problem instance can be easy or hard depending on
the {\it order} in which the variables are assigned values.  This is
because it is possible that the constraints are such that a
particular variable can only take one possible value.  If this variable
is examined early in the search process, the recognition that the value
is highly constrained would permit a large fraction of the search space
to be avoided. Conversely, if this variable is examined late in the
search process, much of the tree might already have been developed,
resulting in relatively little gain. However, the algorithm
described in Sec.~\ref{sect_nested_class} is a naive
algorithm that does {\em not} optimize the order in which the
variables are assigned values. Thus, we can compute the probability $p(i)$
for an average tree having a {\it random} variable ordering.
\par

The simplest way to do this is to consider a {\it lattice} of partial
solutions rather than a {\it tree} of partial solutions, because
a lattice of partial solutions effectively encodes all possible variable
orderings. In particular, the $i$th level of a lattice of partial solutions
represents all possible subsets of $i$ variables out of $\mu$ variables,
assigned values in all possible combinations. Thus, in a lattice there
are ${\mu\choose i} b^i$ nodes at level $i$ rather than the $b^i$ nodes in a
tree. So each level of the lattice encodes the information contained in
${\mu\choose i}$ different trees.
As each constraint involves exactly $k$ variables, and each variable can
be assigned any one of its $b$ allowed values, there are exactly $b^k$
``ground instances'' of each constraint.  Moreover, as each constraint
involves a different combination of $k$ out of a possible $\mu$ variables,
there can be at most ${\mu\choose k}$ constraints. Each ground instance of
a constraint may be ``good'' or ``nogood'', so the number of ground instances
that are ``nogood'', $\xi$, must be such that
$0 \le \xi \le b^k {\mu\choose k}$. If $\xi$ is small the problem
typically has many solutions. If $\xi$ is large
the problem typically has few, or perhaps no, solutions.  The exact
placement of the $\xi$ nogoods is, of course, important in determining the
their ultimate pruning power.
\par

Thus to estimate $p(i)$ in an {\it average} tree, we calculate the
corresponding probability that a node in the lattice (which implicitly
incorporates {\it all} trees) is ``nogood'', conditional on there being $\xi$
``nogoods'' at level $k$.  For a node at level $i$ of the lattice to be
``good'' it must not sit above any of the $\xi$ ``nogoods'' at level $k$.
A node at level $i$ of the lattice sits above ${i\choose k}$ nodes at level
$k$.  Thus, out of a total possible pool of $b^k {\mu\choose k}$ nodes at
level $k$, we must exclude ${i\choose k}$ of them.  However, we can pick
the $\xi$ nogoods from amongst the remaining nodes in any way whatsoever.
Hence the probability that a node is ``good'' at level $i$, given that
there are $\xi$ ``nogoods'' at level $k$, is given by the ratio of the number
of ways to pick the ``nogoods'' such that a particular node at level $i$
is ``good'', to the total number of ways of picking the $\xi$
``nogoods''. As a consequence, 
the probability for a partial solution to be good at level $i$ 
in a tree of height $\mu$ and branching ratio $b$
can be approximated as~\cite{bib_Williams92,bib_Williams93,bib_Williams94}
\begin{equation}  \label{eq_colin_tad}
p(i) = { \displaystyle { b^k {\mu\choose k} - {i\choose k} \choose \xi}   \over
         \displaystyle { b^k {\mu\choose k} \choose \xi}  }
\end{equation}
where $k$ is the size of the constraint (i.e., number of variables
involved in a constraint) and $\xi$ is the number of
``nogood'' ground instances (or number of constraints).
This approximation essentially relies on the assumption
that the partial solutions at a given level are uncorrelated.
\par

Now, we are interested in obtaining an asymptotic expression
for $p(i)$ for large problems, i.e., when the number of variables
$\mu\to\infty$. Recall that to scale a constraint satisfaction
problem up, however, it is not sufficient to increase only $\mu$.
In addition, we ought also to increase the number of constraints
so as to preserve the ``constrainedness-per-variable'', $\beta=\xi/\mu$.
Thus, when we consider scaling our problems up, as we must 
do to assess the asymptotic behavior of the classical and quantum
structured search algorithms, we have $\mu\to\infty$ and scale 
$\xi = \beta \mu$, keeping $\beta$, $b$ and $k$ constant.\footnote{For  
graph coloring, this scaling assumption corresponds to
adding more edges to the graph as we allow the number of nodes to go to
infinity, while simultaneously keeping the average connectivity
(number of edges per node) and the number of colors fixed.}
We now make the assumption
that $\xi \ll  b^k {\mu\choose k}$ and
$\xi \ll b^k {\mu\choose k} - {i\choose k}$, which is justified
in the asymptotic regime. 
Using Stirling formula, we have
\begin{equation}   \label{eq_approxbinom}
{ \displaystyle {M\choose K}\over \displaystyle {N\choose K} }
\simeq { (M-K)^K \over (N-K)^K }
\simeq \left({M\over N}\right)^K
\end{equation}
for large $M$ and $N$,
provided that $K\ll M,N$. This allows us to reexpress
Eq.~(\ref{eq_colin_tad}) as
\begin{equation}
p(i)= \left( 1 - b^{-k} {{i\choose k} \over
                          {\mu\choose k}} \right)^\xi
\end{equation}
Now, assuming that 
$k \ll i$ and $k \ll \mu$, and reusing Eq.~(\ref{eq_approxbinom}), 
we have
\begin{equation}
p(i) = \left( 1 - b^{-k} \left({i \over \mu}\right)^k \right)^\xi
\end{equation}
for large $i$ and $\mu$. Finally, assuming for simplicity that 
$b^k \gg 1$ and $(i/\mu)^k \ll 1$,
we obtain
\begin{equation}   \label{eq_p(i)_app}
p(i)= b ^ {\displaystyle  - \mu \left({\beta\over\beta_c}\right) 
                         \left({i\over \mu}\right)^k }
\end{equation}
where $\beta=\xi/\mu$ measures the difficulty of the problem
and $\beta_c= b^k \log(b)$ is the critical value around which
the problem is the most difficult. 
\par

\section{Average-case complexity of the classical search}
\label{app_classcomplexity}

Plugging Eq.~(\ref{eq_p(i)_app}) into Eq.~(\ref{eq_Tclass}), we obtain
an approximate expression of the classical computation time
needed to solve an average instance with fixed $\beta$
\begin{equation}
T_c(i) = { b^i + b^{\mu-\mu(\beta/\beta_c)(i/\mu)^k} 
       \over b^{\mu-\mu(\beta/\beta_c)} }
\end{equation}
where the denominator is simply the expected number of solutions.
Let us now find the level $i$ where it is optimum to ``cut'' the tree.
The value of $i$ which
minimizes $T_c(i)$ corresponds, for large $\mu$, to the situation
where both terms in the numerator grow with the same power of $b$,
i.e., the solution of the equation $i=\mu-\mu(\beta/\beta_c)(i/\mu)^k$.
Then, one can show that
the computation time approximately scales as
\begin{equation}
T_c  \simeq {2 b^{\alpha \mu} \over b^{\mu-\mu(\beta/\beta_c)}}
\end{equation}
where the scaling coefficient
$\alpha=x$ with $x=i/\mu$, the fraction of the height at which
one cuts the tree, being the solution of Eq.~(\ref{eq_equationforx})
such that $0\le x \le 1$.
For problems of maximum difficulty ($\beta=\beta_c$), i.e., problems
which admit a single solution on average, the classical time scales thus as
\begin{equation}
T_c \simeq {2 d^\alpha \over d^{1-\beta/\beta_c} } = O(d^\alpha)
\end{equation}
for a search space of dimension $d=b^\mu$. This
represents a significant improvement over
a classical search that does not exploit the structure,
i.e., $T_c\sim O(d)$.

\section{Exact expression for the iterated search operator
in terms of Chebyshev polynomials}
\label{app_chebyshev}

The unstructured quantum search algorithm is based
on iterating $n$ times the operator
\begin{equation}  \label{eq_defQ_app}
Q = \left(
\begin{array}{cc}
1-4\,|u|^2 & 2u \\
-2u^*    & 1
\end{array}  \right) 
\end{equation}
where $u \equiv \langle t|U|s\rangle$ is a c-number (with $|u|\le 1$).
The iterated operator can be written exactly as
\begin{equation} \label{eq_Chebyshev}
Q^n  =  (-1)^n  \left(
\begin{array}{cc}
U_{2n}(|u|) & -{\displaystyle u\over\displaystyle |u|} U_{2n-1}(|u|) \\
{\displaystyle u^*\over\displaystyle |u|} U_{2n-1}(|u|) & -U_{2n-2}(|u|)
\end{array}  \right) 
\end{equation}
where $U_n(\cos\theta)=\sin((n+1)\theta)/\sin(\theta)$
is the Chebyshev polynomial of the second kind.
By making use of $U_0(x)=1$, $U_1(x)=2x$, and $U_2(x)=4x^2-1$,
it is easy to check that, for $n=1$, Eq.~(\ref{eq_Chebyshev})
\begin{equation}
Q=-\left(
\begin{array}{cc}
U_2(|u|) & -{\displaystyle u\over\displaystyle |u|} U_1(|u|) \\
{\displaystyle u^*\over\displaystyle |u|} U_1(|u|) & -U_0(|u|)
\end{array}  \right) 
\end{equation}
is indeed consistent with Eq.~(\ref{eq_defQ_app}).
Now, using the recursion formula for Chebyshev polynomials,
\begin{equation}
U_{n+1}(x)-2x U_n(x) + U_{n-1}(x)=0
\end{equation}
we can verify easily that the product of $Q$ and $Q^n$,
as defined by Eq.~(\ref{eq_Chebyshev}),
yields $Q^{n+1}$. Indeed,
\begin{eqnarray}
Q Q^n &=& (-1)^{n+1} \left( \begin{array}{cc}
4|u|^2-1 & -2u \\
2u^* & -1
\end{array}  \right) \;
\left(
\begin{array}{cc}
U_{2n}(|u|) & -{\displaystyle u\over\displaystyle |u|} U_{2n-1}(|u|) \\
{\displaystyle u^*\over\displaystyle |u|} U_{2n-1}(|u|) & -U_{2n-2}(|u|)
\end{array}  \right) 
\nonumber \\
&=& 
(-1)^{n+1}  \left(
\begin{array}{cc}
U_{2n+2}(|u|) & -{\displaystyle u\over\displaystyle |u|} U_{2n+1}(|u|) \\
{\displaystyle u^*\over\displaystyle |u|} U_{2n+1}(|u|) & -U_{2n}(|u|)
\end{array}  \right) 
\nonumber \\
&=& Q^{n+1}
\end{eqnarray}
We can use Eq.~(\ref{eq_Chebyshev}) to calculate the exact amplitude
of the target state $|t\rangle$ after $n$ iterations, that is
\begin{equation}
\langle t|Q^n U |s\rangle
=(-1)^n \left( u\, U_{2n}(|u|) - 
{\displaystyle u\over\displaystyle |u|} \,
U_{2n-1}(|u|) \right)
\end{equation}
Equivalently, we can write
\begin{equation}
\langle t|Q^n U |s\rangle
=(-1)^n \left( u\, T_{2n}(|u|) - 
{\displaystyle u\over\displaystyle |u|} \,
(1-|u|^2) \, U_{2n-1}(|u|) \right)
\end{equation}
by using the recursion formula
\begin{equation}
T_n(x)=U_n(x)-xU_{n-1}(x)
\end{equation}
where $T_n(\cos\theta)=\cos(n\theta)$ is the
Chebyshev polynomial of the first kind.
Note that, at the limit of $|u|\ll 1$, it is easy to show
that $T_{2n}(|u|) \simeq (-1)^n \cos(2n|u|)$
and $U_{2n-1}(|u|) \simeq - (-1)^n \sin(2n|u|)$,
so that we obtain
\begin{equation}  \label{eq_C9}
\langle t|Q^n U |s\rangle
\simeq u \, \cos(2n|u|) + {\displaystyle u\over\displaystyle |u|}
\sin(2n|u|)
\end{equation}
in agreement with Eq.~(\ref{eq_10}).
Thus, the second term in Eq.~(\ref{eq_C9}) mainly contributes to the
amplitude of the target state $|t\rangle$ at the limit of small $|u|$.


\end{document}